\documentclass[aps,prb,reprint,superscriptaddress,letterpaper]{revtex4-1}
\usepackage[english]{babel}
\usepackage{ucs}
\usepackage[utf8x]{inputenc}
\usepackage{graphicx}
\usepackage{bm}
\usepackage{bbm}
\usepackage{empheq}
\usepackage{mathtools}
\newcommand{\bra}[1]{\langle #1|}
\newcommand{\ket}[1]{|#1\rangle}

\newcommand{\mm}[1]{\mathrm{#1}}
\newcommand{\ui}{\mathrm{i}}
\newcommand{\ue}{\mathrm{e}}
\newcommand{\uh}{\mathrm{h}}
\newcommand{\uhu}{\mathrm{h}_{\uparrow}}
\newcommand{\uhd}{\mathrm{h}_{\downarrow}}
\newcommand{\utu}{\mathrm{t}_{\uparrow}}
\newcommand{\utd}{\mathrm{t}_{\downarrow}}

\newcommand{\rv}{\mbox{\boldmath$r$}}
\newcommand{\kv}{\mbox{\boldmath$k$}}
\newcommand{\ev}{\mbox{\boldmath$e$}}
\newcommand{\kvs}{\mbox{\boldmath$\scriptstyle{k}$}}
\newcommand{\dv}{\mbox{\boldmath$d$}}
\newcommand{\Ev}{\mbox{\boldmath$E$}}
\newcommand{\vv}{\mbox{\boldmath$v$}}
\newcommand{\abs}[1]{\left|#1\right|}

\begin{document}

\title{Quantum limit for nuclear spin polarization in semiconductor quantum dots}

\author{Julia Hildmann}
\thanks{Current address: Department of Physics, McGill University, Montreal, Quebec, H3A 2T8, Canada.} 
\affiliation{Department of Physics, University of Konstanz, D-78457 Konstanz, Germany}

\author{Eleftheria Kavousanaki}
\affiliation{Department of Physics, University of Konstanz, D-78457 Konstanz, Germany}
\affiliation{Femtosecond Spectroscopy Unit, Okinawa Institute of Science and Technology, 
Graduate University, Okinawa, 904-0412 Japan}

\author{Guido Burkard}
\affiliation{Department of Physics, University of Konstanz, D-78457 Konstanz, Germany}

\author{Hugo Ribeiro}
\thanks{Current address: Department of Physics, McGill University, Montreal, Quebec, H3A 2T8, Canada.} 
\affiliation{Department of Physics, University of Konstanz, D-78457 Konstanz, Germany}
\affiliation{Department of Physics, University of Basel, Klingelbergstrasse 82, CH-4056
Basel, Switzerland}

\footnotetext{current address}

\begin{abstract}
A recent experiment [E. A. Chekhovich \emph{et al.}, Phys.  Rev. Lett. \textbf{104},
066804 (2010)] has demonstrated that high nuclear spin polarization can be achieved in
self-assembled quantum dots by exploiting an optically forbidden transition between a
heavy hole and a trion state. However, a fully polarized state is not achieved as
expected from a classical rate equation. Here, we theoretically investigate
this problem with the help of a quantum master equation and we demonstrate that a fully
polarized state cannot be achieved due to formation of a nuclear dark state.  Moreover,
we show that the maximal degree of polarization depends on structural properties of the
quantum dot.
\end{abstract}

\maketitle

\section{Introduction}

Initialization~\cite{ono2002}, coherent manipulation~\cite{petta_science2005,
koppens_nature2006, nowack2007, petta2010, ribeiro2013}, and readout of a single spin
confined in a quantum dot have become a common routine. However, and in spite of all
remarkable developed strategies~\cite{kloeffel2013}, a scalable quantum computer based on
spin-qubits~\cite{loss1998} still faces serious challenges.
The main difficulty that has been encountered comes from the unavoidable coupling between
the qubit and the surrounding environment. The time evolution of the qubit becomes
correlated with the dynamics of the environment degrees of freedom. This would not be a
problem in itself if one knew how to control the environment, but in general this cannot
be done, and thus the random character (mixed state) of the environment results in the
decoherence of the qubit~\cite{chirolli2008}.

In quantum dots made out of III-V materials, the hyperfine interaction of a single
electron with a large number of nuclear spins ($10^4$ -- $10^6$) is the main source of
decoherence~\cite{coish_statsolidi2009,petta_science2005,koppens_nature2006,khaetskii_prl2002,
merkulov_prb2002}. However, through an extensive effort aiming at prolonging the spin
coherence in quantum dots nanostructure, several approaches have been put forward to minimize or
even cancel the effects due to nuclear spin induced dynamics. Dynamical decoupling
techniques, such as Hahn echo~\cite{hahn1950} or Carr-Purcell~\cite{carr1954}, allow a
refocusing of the qubit phase by eliminating the low-frequency components of the nuclear
spin bath fluctuations. These methods have demonstrated that it is possible to extend the
inhomogeneous dephasing time $T_2^* \sim\,10\,\mm{ns}$~\cite{petta_science2005,
koppens_nature2006,coish_prb2004, coish2005, xu2009} up to the dephasing time
$T_2\sim\,3\,\mm{\mu
s}$~\cite{petta_science2005,koppens_nature2006,greilich2009,clark2009,press2010}, which
corresponds to the limit imposed by nuclear spin diffusion~\cite{coish_prb2004,
witzel2006}. A more recent experiment in gate defined double quantum dots has even
revealed $T_2 \simeq 200\,\mm{\mu s}$~\cite{bluhm2011}.

Another possible route consists in polarizing the nuclear spins. However, a substantial
degree of polarization (close to 100\%) is needed~\cite{coish_prb2004} to increase
coherence times. Highly polarized nuclear states are also desirable for other useful
tasks in quantum information. Ultimately they can be used as a quantum memory to store
the coherent state of the electron spin~\cite{taylor_prl2003, schwager_prb2010}.
Nuclear spins represent an attractive system for this purpose, since the nuclear
polarization can persist for minutes in the dark~\cite{maletinsky2007,nikolaenko2009} (in
absence of an electron in the dot). Despite huge breakthroughs in coherent control of
nuclear spin polarization \cite{gammon_science1997, carter_prl2009, makhonin_natmat2011},
switching its direction \cite{latta_nature2009}, observing reversal behavior
\cite{tartakovskii_prl2007} and controlling only certain group of nuclear spins
\cite{makhonin_prb2010}, a close to 100\% polarized nuclear state has yet to be reported. 

A new experimental method relying on spin-forbidden transitions between heavy
holes and trions (positively charged excitons consisting of two heavy holes in a singlet
state and an electron) was believed to be capable of fully polarizing the nuclear spin
bath. This expectation was based on a rate equation describing the pumping mechanism
which was predicting a fully polarized nuclear state~\cite{chekhovich_prl2010}. Although
the reached polarization was one of the highest until now reported, $\sim
65$\%~\cite{chekhovich_prl2010}, it is still below the threshold
required for reliable quantum information processing.

The inability to reach a maximally polarized nuclear state shows that our understanding
of the hyperfine mediated dynamics is still incomplete. In this article, we develop a
model of optical nuclear spin polarization, as studied experimentally
in Ref.~[\onlinecite{chekhovich_prl2010}]. Our theory 
goes beyond the commonly used description of the nuclear spins as a stochastic magnetic
field~\cite{khaetskii_prl2002, coish2005}.  We take into account the quantum nature of the
nuclear spins and use a fully quantum mechanical master equation describing the
joint time evolution of the electronic and nuclear degrees of freedom. In particular, we
show that the pumping saturation is a consequence of the collective nuclear spin
dynamics. 
By studying both cases of homogeneous and inhomogeneous hyperfine coupling constants, we
show that the simpler case of homogeneous coupling qualitatively describes all physical
phenomena. The inhomogeneous case, since more close to experimental conditions, provides
quantitative agreement with the experiment. We also investigate in more detail the
variation in the degree of maximal possible polarization depending on the distribution of
the electron wave function inside of the quantum dot relative to the lattice using a
shell model~\cite{tsyplyatyev_prl2011}.

\section{System Hamiltonian}

We start with the following Hamiltonian,
\begin{equation}
	H(t) = H_0 + H_{\mm{L}}(t) + H_{\mm{HF}},
\label{eq:Htotal}
\end{equation}
where $H_0$ describes the electronic system, $H_{\mm{L}}(t)$ its interaction
with the laser field, and $H_{\mm{HF}}$ the effective hyperfine
interaction with the nuclear spins.

\begin{figure}
\centering
\includegraphics[width=0.45\textwidth]{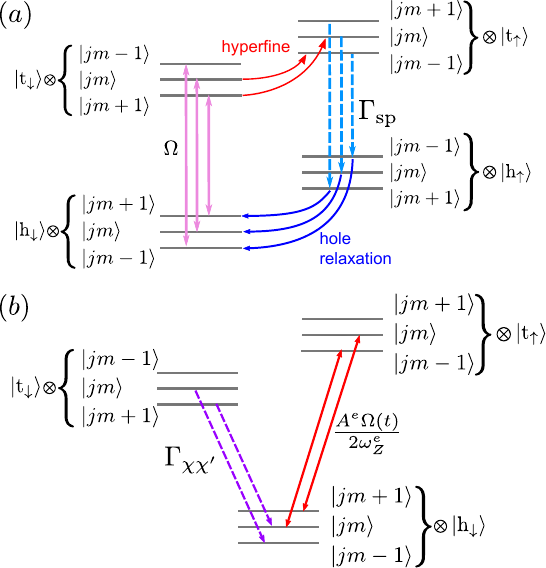}
\caption{(a) Level scheme for electronic and nuclear states. The hyperfine sublevels are
denoted with their total nuclear spin quantum numbers $j$ and $m$. The trion state
$\ket{\utd}$ with angular momentum (along $z$) $M=-1/2$ is pumped by the laser
light with Rabi frequency $\Omega$ from the heavy hole state $\ket{\uhd}$ with
$M=-3/2$. The hyperfine interaction couples the trion state $\ket{\utd}$ and
$\ket{\utu}$ ($M=1/2$) and changes the $m$ quantum number of the nuclear system.
The trion states can relax by spontaneous emission with the rate $\Gamma_{\mm{sp}}$.  (b)
Reduced level scheme with mechanisms for nuclear polarization in the trion - heavy hole
system by spin-forbidden relaxation with rate $\Gamma_{\chi \chi'}$ from $\ket{\utd}$, or
by spin-forbidden optical transitions between $\ket{\uhd}$ and $\ket{\utu}$.}
\label{fig:level_scheme}
\end{figure}

The electronic system consists of four levels: heavy hole with spin up (hole-up, $\ket{3/2,
3/2} \equiv \ket{\uhu}$), heavy hole with spin down (hole-down, $\ket{3/2, -3/2} \equiv
\ket{\mm{h}_\downarrow}$), trion with electron spin up (trion-up, $\ket{1/2, 1/2} \equiv
\ket{\mm{t}_\uparrow}$), and trion with electron spin down (trion-down, $\ket{1/2, -1/2}
\equiv \ket{\mm{t}_\downarrow}$) [c.f.  Fig.~\ref{fig:level_scheme}(a)]. The Hamiltonian
$H_0$ of these four states in the presence of an external homogeneous magnetic field is
given by
\begin{equation}
	H_0=E_{\mm{t}}\tau_{\ue}+ \omega_Z^{\ue} S^{\ue}_z + \omega_Z^{\uh} S^{\uh}_z.
\label{eq:H0elec}
\end{equation}
Here $E_{\mm{t}}$ is the energy needed to excite a heavy hole to a trion and $\tau_{\ue}$
represents the projection operator onto the trion spin states, $\tau_{\ue} =
\ket{\utd}\bra{\utd} + \ket{\utu}\bra{\utu}$. The trion (heavy hole) Zeeman splitting is
given by $ \omega_Z^{\ue}=g_{\ue}\mu_B B_z\,(\omega_Z^{\uh}=g_{\uh}\mu_B B_z)$, where
$g_{\ue}\,(g_{\uh})$ is the electron (heavy hole) Landé $g$-factor, $\mu_B$ is the Bohr
magneton and $B_z$ is the external magnetic field chosen along the growth axis of the
quantum dot. We use $B_z=2.5\,\mm{T}$ and $g_{\ue} =1.5$ (measured in
Ref.~\onlinecite{chekhovich_prl2010}).  $S_z^{\ue}$ is the trion spin operator and
$S_z^{\uh}$ is the pseudo-spin operator for heavy hole spin states along the direction of
the magnetic field. 

The laser Hamiltonian $H_{\mm{L}}(t)$ describes the  left circularly polarized laser field
that pumps the transition between heavy hole-down $\ket{\uhd}$ and trion-down
$\ket{\utd}$ ($M=-1/2$) states,
\begin{equation}
	H_{\rm L}(t)=\hbar\Omega\,\left(\ue^{-\ui\omega_{\mm{L}} t}\ket{\utd}\bra{\uhd}
	+ \ue^{\ui \omega_{\mm{L}} t}\ket{\uhd}\bra{\utd}\right).
	\label{eq:Hlaser}
\end{equation}
Here $\omega_{\mm{L}}$ is the laser frequency and $\Omega$ is the Rabi frequency. In our
calculations we use $\Omega=20\,\mm{GHz}$.  In principle, the Rabi frequency is a
function of time. However, since the pumping time is much larger than the
characteristic time needed to switch the laser on and off, $t_{\mm{pump}} \gg
\tau_{\mm{on/off}}$, we assume a constant intensity of the laser light during the
whole pumping cycle.

The hyperfine Hamiltonian includes the contributions from both the electron and the heavy
hole. It is described by the effective Hamiltonian,
\begin{equation}
	H_{\mm{HF}} = \sum_{k=1}^N \left[ \frac{1}{2} A_k^{\ue} \left( 2 S_z^{\ue} I_z^k + S_+^{\ue} I_-^k +
	S_-^{\ue} I_+^k\right) + A_k^{\uh} S_z^{\uh} I_z^k\right], 
	\label{eq:Hhf}
\end{equation}
where the coupling to the hole states is strongly anisotropic \cite{fischer_prb2008}.
Here, the sum runs over all $N$ nuclei within the quantum dot. The operator $I_z^k$
describes the $z$ component of the $k$th nuclear spin. In Eq.~\eqref{eq:Hhf}, we have
introduced the spin ladder operators defined as $S_\pm^{\ue} = S_x^{\ue} \pm \ui
S_y^{\ue}$ and $I_\pm^k = I_x^k \pm \ui I_y^k$.  The hyperfine coupling constants with
the $k$th nucleus are given by $A_k^{\ue} = v_{k}^{\ue} \nu_0 \left|\psi^{\ue}
(\rv_k)\right|^2$ and $A_k^{\uh} = v_{k}^{\uh} \nu_0 \left|\psi^{\uh}(\rv_k)\right|^2$,
where $v_{k}^{\ue (\uh)}$ is the hyperfine coupling strength of the electron spin (heavy
hole), $\nu_0$ the volume of a unit cell, and $\psi^{\ue (\uh)}(\rv_k)$ is the envelope
wave function of the electron (heavy hole).	

The homogeneous approximation of the Hamiltonian~\eqref{eq:Hhf} is performed by
replacing the position-dependent coupling constant by $A^{\ue}/N$, where  $A^{\ue}$ is
the average hyperfine coupling constant (for InP quantum dots, $A^{\ue}=110\,\mm{\mu
eV}$~\cite{chekhovich_prl2010}).  The interaction strength between a heavy hole and the
$k$th nuclear spin is given by $A_k^{\uh}$. It differs from the electron hyperfine constant
due to a different type of wave function. It was found theoretically and confirmed
experimentally, that $A^{\ue}\approx -0.11 A^{\ue}$ \cite{fischer_prb2008, chekhovich_prl2011,
fallahi_prl2010}.

We omit the transverse terms of the effective heavy hole hyperfine interaction, which can
contribute to nuclear spin polarization~\cite{fischer_prb2008}. The coupling constants
for the longitudinal and transverse hyperfine terms of the heavy hole are different due
to the anisotropic character of the interaction. This leads to a transverse hyperfine
coupling constant which is approximately two orders of magnitude smaller than the
longitudinal one, $\abs{A_{\perp}^{\uh}} < 0.06 \abs{A_z^{\uh}}$~\cite{fischer_prb2008}.
In addition,
the large Zeeman energy ($B=2.5\,\mm{T}$) renders hyperfine assisted relaxation of heavy
holes small compared to other physical mechanisms playing a role in the polarization of
nuclear spins~\cite{fras2012}.

In the following, we split the hyperfine Hamiltonian into longitudinal and transverse
contributions. The longitudinal term,
\begin{equation}
	H_{\mm{HF}}^z = \sum_{k=1}^N \left( A_k^{\ue} S_z^{\ue} I_z^k +  A_k^{\uh}
	S_z^{\uh} I_z^k\right),
	\label{eq:Hhfl}
\end{equation}
only produces a spin-dependent energy shift (Overhauser shift) of the electronic states, while the
transverse part,
\begin{equation}
	H_{\mm{HF}}^{\perp} = \frac{1}{2} \sum_{k=1}^N A_k^{\ue} \left(S_+^{\ue} I_-^k +
	S_-^{\ue} I_+^k \right)
	\label{eq:Hhft}
\end{equation}
provides the mechanism for polarizing the nuclear spins by transferring magnetic moment
from the electron spin to the nuclear spin ensemble.

The time dependence of Hamiltonian~\eqref{eq:Htotal} can be removed by performing a
canonical transformation,
\begin{equation}
	H \to H'={\rm e}^{\ui \xi t/\hbar}(H-\xi){\rm e}^{-\ui\xi t/\hbar}.
	\label{eq:canonical}
\end{equation}
For our problem, we have
\begin{equation}
	\xi=\left( E_{\mm{t}} - \frac{\hbar}{2}\omega_Z^{\ue} -
	\frac{\hbar\Delta}{2}\right) \tau_{\mm{e}} - \hbar\left(\omega_Z^{\uh} +
	\Delta\right)\tau_{\uh},
	\label{eq:xi}
\end{equation}
where $\tau_{\uh} = \ket{\uhd}\bra{\uhd} + \ket{\uhu}\bra{\uhu}$ is the projection
operator onto the heavy-hole spin states. The detuning of the laser frequency from the
heavy hole down to trion down, $\ket{\uhd} \rightarrow |\utd\rangle$, transition
energy is given by $\Delta =
E_{\mm{t}}/\hbar+\tfrac{1}{2}(\omega_Z^{\uh}-\omega_Z^{\ue})-\omega_{\mm{L}}$. The
transformation only acts non-trivially on $H_0$ [Eq.~\eqref{eq:H0elec}], and
$H_{\mm{L}}(t)$ [Eq.~\eqref{eq:Hlaser}]. We find
\begin{equation}
	H_0 \to H_0'=\hbar \omega_Z^{\ue} \ket{\utu}\bra{\utu}
		    + \frac{\hbar\Delta}{2} \tau_{\ue} 
		    + \hbar\omega_Z^{\uh} \ket{\uhu}\bra{\uhu} 
		    - \frac{\hbar\Delta}{2}\tau_{\uh},
	\label{eq:H0p}
\end{equation}
and
\begin{equation}
	H_{\mm{L}}(t) \to H_{\mm{L}}'=\hbar\Omega\left(\ket{\utd}\bra{\uhd} +
	\ket{\uhd}\bra{\utd}\right), 
	\label{eq:Hlaserp}
\end{equation}
after performing the rotating-wave approximation on $H_{\mm{L}}'$.

The Hamiltonian defined in Eq.~\eqref{eq:Htotal} becomes then,
\begin{equation}
	H'=H_0'+ H_{\rm L}'+ H_{\mm{HF}}^z + H_{\mm{HF}}^{\perp}.
	\label{eq:HtotalRWA}
\end{equation}

We further eliminate the hyperfine spin-flip terms from Eq.~\eqref{eq:HtotalRWA} by
applying a Schrieffer-Wolff transformation~\cite{latta_nature2009, issler_prl2010,
bravyi_annphys2011}
\begin{equation}
	H' \to \tilde{H} = \ue^S H' \ue^{-S}= \sum_{j=0}^{\infty}
	\frac{\left[S,H'\right]^{(j)}}{j!},
	\label{eq:SW}
\end{equation}
where we have used the recursive definition
\begin{equation}
	\begin{aligned}
		\left[S,H'\right]^{(0)} &= H' , \\ 
	        \left[S,H'\right]^{(1)} &= \left[S,H'\right] , \\ 
		\left[S,H'\right]^{(j)}  &= \left[S, \left[S,H'\right]^{(j-1)}\right].
	\end{aligned}
	\label{eq:defad}
\end{equation}

By applying the Schrieffer-Wolf transformation as defined in Eq.~\eqref{eq:SW} with
\begin{equation}
	S=\sum_{k=1}^N\frac{A_k^{\ue}}{2\omega_Z^{\ue}}\left(I_-^k \ket{\utu}\bra{\utd} -
	I_+^k \ket{\utd}\bra{\utu}\right), 
	\label{eq:SSW}
\end{equation}
we obtain an effective Hamiltonian with hyperfine interaction assisted spin-forbidden
optical transitions:
\begin{equation}
	\tilde{H}=H_0'+ H_{\rm L}'+ H_{\mm{HF}}^z +
	\sum_{k=1}^N\frac{\hbar A_k^{\ue}\Omega}{2\omega_Z^{\ue}}\left(I_-^k 
	\ket{\utu}\bra{\uhd} + I_+^k \ket{\uhd}\bra{\utu}\right).
\label{eq:Hfinal}
\end{equation}
In the Hamiltonian~\eqref{eq:Hfinal}, we only include terms of the Schrieffer-Wolff
transformation up to the first order in $A^{\ue}_k$. Higher-order terms describe, e.g.,
second order processes such as extrinsic nuclear-nuclear spin interactions assisted by
two virtual electron spin flips~\cite{yao_prb2006, latta_prl2011,klauser_2006}, which are of
little interest here.

The effective Hamiltonian defined in Eq.~\eqref{eq:Hfinal} gives an intuitive picture of
the optical pumping mechanism from heavy hole down to trion up [c.f.
Fig.~\ref{fig:level_scheme}(b)]. When the laser frequency is on resonance with the
transition $\ket{\uhd} \to \ket{\utu}$, i.e., $\Delta = -\omega_Z^{\ue}$, simultaneous
absorption of a photon and transfer of angular momentum from the electron spin to the
nuclear spin bath takes place. However, the coherent dynamics alone cannot explain the
build-up of nuclear polarization.  To correctly describe the pumping cycles, we need to
include the spontaneous emission of the trion state. In this scenario, the quantum dot is
initialized in the state $\ket{\uhd}$, optically pumped to the state $\ket{\utu}$,
simultaneously transferring the angular momentum of the electron spin to the nuclear
bath, and the trion-up state decays by spontaneous emission to the state $\ket{\uhu}$
faster than it can be optically pumped back to $\ket{\uhd}$.  The heavy hole up relaxes
then via spin orbit to the initial heavy hole down~\cite{climente2013} and another
pumping cycle can start again. To describe the evolution of the system in presence of
dissipation, we rely on the Lindblad master equation~\cite{lindblad1976,
breuer_petruccione}.

\section{Lindblad Master Equation}

The Lindblad master equation for the density matrix $\rho$ of the combined electronic and
nuclear spin system is given by
\begin{equation}
	\dot{\rho} = -\frac{\ui}{\hbar}\left[\tilde{H},\rho\right] +
	\frac{1}{2}\sum_{j=1}^{d^2 -1} \left(\left[L_j \rho,
	L_j^{\dag}\right] + \left[L_j, \rho
	L_j^{\dag}\right]\right),
	\label{eq:mastereqL}
\end{equation}
where the Lindblad operators $L_j$ describe different dissipation
processes~\cite{lindblad1976, breuer_petruccione}, and $d$ is the dimension of the
Hilbert space. Here, we only describe dissipative processes in the low-dimensional
electronic system and therefore get by with a small number of Lindblad operators.

As mentioned earlier, a key dissipation process to explain the polarization dynamics is
the spontaneous emission. Here, we take into account the spontaneous emission of a photon
from the trion-down state to the corresponding hole state. This process is described by
$L_1 = \sqrt{\Gamma_{\mm{sp}}} \ket{\uhd}\bra{\utd}$, with $\Gamma_{\mm{sp}}=6
\,\mm{GHz}$. In principle, there is a similar process for the other trion and hole states
and a process that describes the relaxation of the heavy hole up to heavy hole down [cf.
Fig.~\ref{fig:level_scheme}(a)]. However, since the heavy hole up does not play any major
role in the polarization dynamics, we are going to consider a direct decay mechanism from
the trion up to the heavy hole down, $L_2 = \sqrt{\Gamma_{\mm{sp}}}\ket{\uhd}\bra{\utu}$.
This is justified if the relaxation from heavy hole up to heavy hole down is not the
bottleneck of the pumping cycle, i.e., the hole relaxation rate is considerably higher
than the nuclear spin pumping rate. The latter has been demonstrated by Checkhovich
\emph{et~al}. in Ref.~[\onlinecite{chekhovich_prl2010}]. Thus, we assume a heavy-hole
relaxation rate that fulfills $1/\Gamma_{\mm{sp}} + 1/\Gamma_{\uhu \to \uhd} \approx
1/\Gamma_{\mm{sp}}$. Finally, we note that heavy-hole relaxation rates as short as
$10\,\mm{ps}$ were reported in Ref.~[\onlinecite{climente2013}]. This allows us to reduce the
dimension of the electronic Hilbert subspace by omitting the heavy hole-up. 

In addition to these two relaxation mechanisms, we include an additional process to
describe the experimentally observed nuclear polarization when the laser frequency is on
resonance with the transition $\ket{\uhd} \to \ket{\utd}$, i.e. $\Delta = 0$. The
mechanism leading to polarization in this case is substantially different from what
happens when $\Delta = -\omega_Z^{\ue}$. The heavy hole down is optically excited to the
trion-down state. At this point, there are two different relaxation paths: the trion down
can either relax back to the heavy hole down by both spontaneous or stimulated emission or it
can relax to the heavy hole up by transferring angular momentum to the nuclear bath. We
describe the latter mechanism with the Lindblad operators, $L_{\chi \chi'} = \sqrt{\Gamma_{\chi
\chi'}}\ket{\uhu\,\chi}\bra{\utd\, \chi'} + \sqrt{\Gamma_{\uhu \to \uhd}}\ket{\uhd
\chi}\bra{\uhu \chi}$, where $\chi=j_1m_1 \cdots j_nm_n$ labels collective angular
momentum states of nuclear spins, which have been arranged into $n$ groups according to
their hyperfine coupling strength $A_i^{\ue}$ with the electronic spin. Since
$\Gamma_{\uhu \to \uhd} \gg \Gamma_{\mm{sp}}$ and $\Gamma_{\chi \chi'} \propto
\Gamma_{\mm{sp}}$,  we can approximate $L_{\chi \chi'}$ to $L_{\chi \chi'}\simeq
\sqrt{\Gamma_{\chi \chi'}}\ket{\uhd\,\chi}\bra{\utd\, \chi'}$. The rates $\Gamma_{\chi
\chi'}$ are calculated in the following with help of Fermi's golden rule.

\subsection{Forbidden relaxation rate}

We describe the interaction of the trion-heavy-hole system with the radiation field
with Hamiltonian $H_{\rm rad}$.  We have, in the dipole approximation~\cite{scullyQO},
\begin{equation}
	H_{\mm{rad}} = - \dv\cdot \Ev,
	\label{eq:afint}
\end{equation}
where $\dv$ is the dipole operator of the quantum dot states and $\Ev$ is the quantized electric field,
\begin{equation}
	\Ev = \sum_{\kvs,\, \lambda} \sqrt{\frac{\hbar \omega_{k}}{2 \varepsilon_0
	V}} \ev_{\kvs,\, \lambda}\left(a^\dag_{\kvs,\, \lambda} + a_{\kvs,\,
	\lambda}\right),
	\label{eq:QEfield}
\end{equation}
with $\varepsilon_0$ the vacuum permittivity. We have decomposed the field confined in a
box of volume $V$ into Fourier modes with periodic boundary conditions. Each mode is
associated with a wave vector $\kv$, two transverse polarization vectors $\ev_{\kvs,\,
\lambda}$, and frequency $\omega_{k}$.  Furthermore, we have introduced the annihilation
and creation operators $a_{\kvs,\, \lambda}$ and $a^\dag_{\kvs,\, \lambda}$ of a photon
with wave vector $\kv$ and polarization $\ev_{\kvs,\, \lambda}$.

We will use $H_{\mm{rad}}$ as a perturbation and apply Fermi's golden rule to compute the
rate $\Gamma_{\chi \chi'}$. In order to do so, we first need to find the eigenstates of
$H_0 + H_{\rm HF}^z + H_{\rm{HF}}^\perp$. This becomes particularly arduous due to the hyperfine
interaction. To ease our task, we can instead use approximate eigenstates found with
perturbation theory. This is possible thanks to the large Zeeman splitting of the
electronic states.

We use $H_0 + H_{\mm{HF}}^z$ as unperturbed Hamiltonian and $H_{\mm{HF}}^{\perp}$ as
perturbation. The eigenstates of the unperturbed Hamiltonian can be written as
$\ket{\psi^{(0)}_{e\,\chi}} = \ket{e\, \chi}$, where $\ket{e}$ labels trion and heavy
hole states. Using first order perturbation theory, we find the corrections for the
states $\ket{\uhu\,\chi}$ and $\ket{\utd\,\chi}$, which respectively, read as
\begin{eqnarray}
	\ket{\psi^{(1)}_{\uhu\,\chi}}&=&0;\\
	\ket{\psi^{(1)}_{\utd\,\chi}}&=&\frac{1}{2}\sum_k\frac{A^{\ue}_k \sqrt{j_k(j_k + 1)-m_k(m_k - 1)}}
	{E^{(0)}_{\utd\,\chi}-E^{(0)}_{\utu\,\chi_k}}\times\nonumber\\
	&& \ket{\utu\,\chi_k},
\end{eqnarray}
with $\chi_k = j_1 m_1 \cdots j_k m_k -1 \cdots j_n m_n$.  Here
$E^{(0)}_{\utd\,\chi}=E_{\mm{t}}-\omega^{\ue}_{\mm{Z}}/2-\sum_i A^{\ue}_i m_i/2$ and
$E^{(0)}_{\utu\,\chi_k} = E_{\mm{t}} + \omega^{\ue}_{\mm{Z}}/2 + \sum_i A^{\ue}_i m_i/2 -
A^\ue_k/2$. The first-order correction to the energy is identically zero for all states,
$E^{(1)}_{e\,\chi} = 0$.

The transition rate $\Gamma_{\chi \chi'}$ is then given by 
\begin{equation}
	\Gamma_{\chi \chi'}= \frac{2\pi}{\hbar} \abs{\bra{\mm{f}} H_{\mm{rad}}\ket{\mm{i}}}^2\rho(E_{\mm{i}} - E_{\mm{f}}),
\label{eq:fermi_rule}
\end{equation}
with $\ket{\mm{f}}= \ket{\psi_{\uhu\,\chi'},\,1}$ and $\ket{\mm{i}}=
\ket{\psi_{\utd\,\chi},\,0}$. We denote the state of the radiation field by $\ket{0}$ (no
photon) and $\ket{1}$ (one photon emitted). The energies of the initial and final state are
$E_{\mm{i}} = E_{\mm{t}} - \omega^{\ue}_{\mm{Z}}/2$ and $E_{\mm{f}}=\omega^{\uh}_{\mm{Z}}/2 +
\omega_{\gamma}$, where $\omega_{\gamma}$ is the energy of the emitted photon.

The evaluation of Eq.~\eqref{eq:fermi_rule} yields
\begin{equation}
	\Gamma_{\chi \chi'}  = \frac{\Gamma_{\mm{sp}}}{4} \abs{\sum_k\frac{A^{\ue}_k\sqrt{j_k(j_k+1)-m_k(m_k-1)}}
	{\omega_{\mm{Z}}^{\ue}+\sum_l A_l^{\ue} m_l - \frac{A^\ue_k}{2}}}^2 \delta_{\chi, \chi'_k},
\label{eq:fr_rate}
\end{equation}
where we have used~\cite{scullyQO, breuer_petruccione},
\begin{equation}
	\Gamma_{\mm{sp}}=\frac{2\pi}{\hbar}
	\abs{\bra{\uhu\,1}H_{\mm{rad}}\ket{\utu\,0}}^2\rho(E_{\mm{i}}-E_{\mm{f}}),
	\label{eq:gsp}
\end{equation}
with $\Gamma_{\mm{sp}}$ the spontaneous emission rate of the trion state.

We assume the spontaneous relaxation rate of the trion-up and -down states to be the
same. The spontaneous emission follows a cubic dependence on the energy difference
between the initial and final states, $\Gamma_{\mm{sp}} \propto \omega_{\mm{fi}}^3$.
Having this in mind and noticing that both Zeeman splittings are four orders of magnitude
smaller than the required energy to create a trion, $\hbar \omega^{\ue
(\uh)}_{\mm{Z}}/E_{\mm{t}} \simeq 10^{-4}$, it is perfectly reasonable to assume that
both spontaneous emission rates are nearly identical.

Another process that could lead to nuclear spin polarization is hyperfine-mediated phonon spin
flips. However, from the experimental data presented in
Ref.~[\onlinecite{chekhovich_prl2010}], there is no evidence that such processes play an
important role in the dynamics. Both the absence of polaronic sidebands in the
photoluminescence spectra of the quantum dot and the absence of side peaks in the
measurement of the nuclear spin polarization seem to indicate very weak phonon coupling.
One would indeed expect to see polarization side peaks (both sides of the main peak) if
the emission or absorption of a phonon would assist the allowed (forbidden) transition
when the laser frequency is detuned off resonance.

\subsection{Solutions of the master equation}

Since the Hamiltonian $\tilde{H}$ [Eq.~\eqref{eq:Hfinal}] is time independent, the master
equation~\eqref{eq:mastereqL} is a system of homogeneous differential equations of
first-order. Using a superoperator formalism, we can rewrite Eq.~\eqref{eq:mastereqL} as
\begin{equation}
	\dot{\rho}(t)=\mathcal{L}\rho(t).
	\label{eq:masterEqSO}
\end{equation}
Interpreting the above equation as a vector equation, we can write the solution for $\rho
(t)$ as
\begin{equation}
 	\rho(t)=\sum_i c_i\,\vv_i\, e^{\lambda_i t},
	\label{eq:solL}
\end{equation}
where $\lambda$ and $\vv$ are the eigenvalues and eigenvectors of $\mathcal{L}$, respectively. The
coefficients $c_i$ can be found from the initial conditions, 
\begin{equation}
	\rho(0)=\sum_i c_i \vv_i.
	\label{eq:ci}
\end{equation}

For the initial conditions of the total density matrix we apply the sudden approximation
\cite{coish_prb2004}. We assume that at times $t<0$, the electronic ($\rho_{\ue}$) and
nuclear ($\rho_{\mm{nuc}}$) density matrices are uncorrelated and for $t=0$ the state of the
total system, $\rho$, is given by $\rho(0)=\rho_{\ue}(0)\otimes\rho_{\mm{nuc}}(0)$.
The initial state for the electronic system is,
\begin{equation}
	\rho_{\ue}(t=0) = \ket{\uhd}\bra{\uhd}.
	\label{eq:rhoeinit}
\end{equation}
The initial nuclear state is assumed to be a fully mixed state. This assumption is
justified by the fact that under normal experimental conditions the thermal energy is
much larger than the nuclear Zeeman, $k_{\mm{B}} T \gg E_{\mm{Z}}^{\mm{nuc}}$. Thus, it
is reasonable to assume a fully unpolarized nuclear state,
\begin{equation}
	\rho_{\mm{nuc}} = \sum_{\chi} p(\chi) \ket{\chi}\bra{\chi}.
	\label{eq:rhoninit}
\end{equation}
By assuming the nuclear spins to be spin-$1/2$ and considering the case of
a Dicke state $\ket{j m}$, we derive the probability distribution $p(\chi)$. Since we can
write the probability for a Dicke state as the probability to find a given $j$ times the
conditional probability of finding $m$ knowing $j$, $p(j,m) = p_j (j) p_m (m|j)$, our
task is reduced to find the degeneracy $g(j)$ of the quantum number $j$. We have $p(j,m)
= g(j) p_m (m|j)/\mm{dim}(\mathcal{H})$, where $\mm{dim}(\mathcal{H}) = 2^N$ is the total
number of nuclear spin states. For a thermal state the distribution of $m$ for a
given $j$ is uniform,
\begin{equation}
	p_m (m|j) = \frac{[\Theta(j+m)-\Theta(j-m)]}{2j + 1},
	\label{eq:pmj}
\end{equation}
where $\Theta(x)$ is the Heaviside step function. The degeneracy $g(j)$  can be
found following the method of Ref.~[\onlinecite{mandel_wolf}], we find
\begin{equation}
	g(j) = \frac{(2j + 1)^2 N!}{\left(\frac{N}{2} + j + 1\right)!\left(\frac{N}{2}
	-j\right)!}.
	\label{eq:degj}
\end{equation}
As a simple and intuitive example consider the case of two spins ($N=2$). For this case,
we can explicitly construct the four Dicke states $\{ \ket{0,0}, \ket{1,-1}, \ket{1,0},
\ket{1,1}\}$, which are the well-known singlet ($j=0$) and triplet states ($j=1$).
Equation~\eqref{eq:degj} for $j=0$ and $1$ yields, respectively, $g(0)=1$ and $g(1)=3$.
Combining the previously derived results, we arrive at
\begin{equation}
	p(j, m)= \frac{(2j+1) N! [\Theta(j+m)-\Theta(j-m)]}{\left(\frac{N}{2} + j +
	1\right)!\left(\frac{N}{2}
        -j\right)!2^N}. 
\label{eq:prob}
\end{equation}

This result is straightforwardly generalized for the case of a state $\ket{\chi}$,
\begin{equation}
	p(\chi) = \prod_{i=1}^n p(j_i, m_i),
	\label{eq:probinh}
\end{equation}
with $N$ in Eq.~\eqref{eq:prob} being replaced by $N_i$, i.e. the number of nuclear spins
in group $i$. Finally, we arrive at the expression of the initial density matrix,
\begin{equation}
	\rho(0) = \sum_{\chi} p(\chi) \ket{\uhd \chi}\bra{\uhd \chi}.
	\label{eq:rhoinit}
\end{equation}

\section{Results}

We calculate the nuclear spin polarization as a function of the pumping time
$t_{\mm{pump}}$ and as a function of laser detuning $\Delta$ for a fixed pumping time.
The polarization $P$ is calculated according to
\begin{equation}
	P(t) = \sum_k A_k^{\ue} \langle I^z_k (t) \rangle,
	\label{eq:polarization}
\end{equation}
with $\langle I^z_k(t)\rangle= \mathrm{Tr}\,[I^z_k\rho_{\mm{nuc}} (t)]$ and
$\rho_{\mm{nuc}}(t)=\mathrm{Tr}_{\rm e} [\rho(t)]$ is obtained by taking the partial trace over
the electronic states. This definition corresponds to the experimental procedure that is
employed to measure the nuclear spin magnetization, which is done by measuring the shift
of the electronic Zeeman splitting and interpreting it as an effective magnetic field,
$B_{\mm{nuc}}=\sum_k A_k^{\rm e} \langle I_k^z \rangle/ g^* \mu_{\mm{B}}$.

\subsection{Homogeneous hyperfine coupling}
The simplest way to solve Eq.~\eqref{eq:masterEqSO} is to assume a homogeneous hyperfine
coupling constant, $A_k^{\ue} = A^{\ue}/N$ ($A^{\ue} =110\,\mm{\mu eV}$ for InP quantum
dots~\cite{chekhovich_prl2010}). This model corresponds to the case in which the
electronic envelope wave function in the quantum dot is a plane wave. It is often
referred to as ``box'' model~\cite{kozlov_arxiv2008, petrov_prb2009}. In this case, a
state $\ket{\chi}$ reduces to a Dicke state $\ket{j m}$.
In this basis, the density matrix is block-diagonal (each block corresponding to a fixed
$j$), which allows us to compute the time evolution separately for each block. This is a
direct consequence of the lack of transitions between different $j$ states. 
\begin{figure}[t!]
\centering
\includegraphics[width=1\columnwidth]{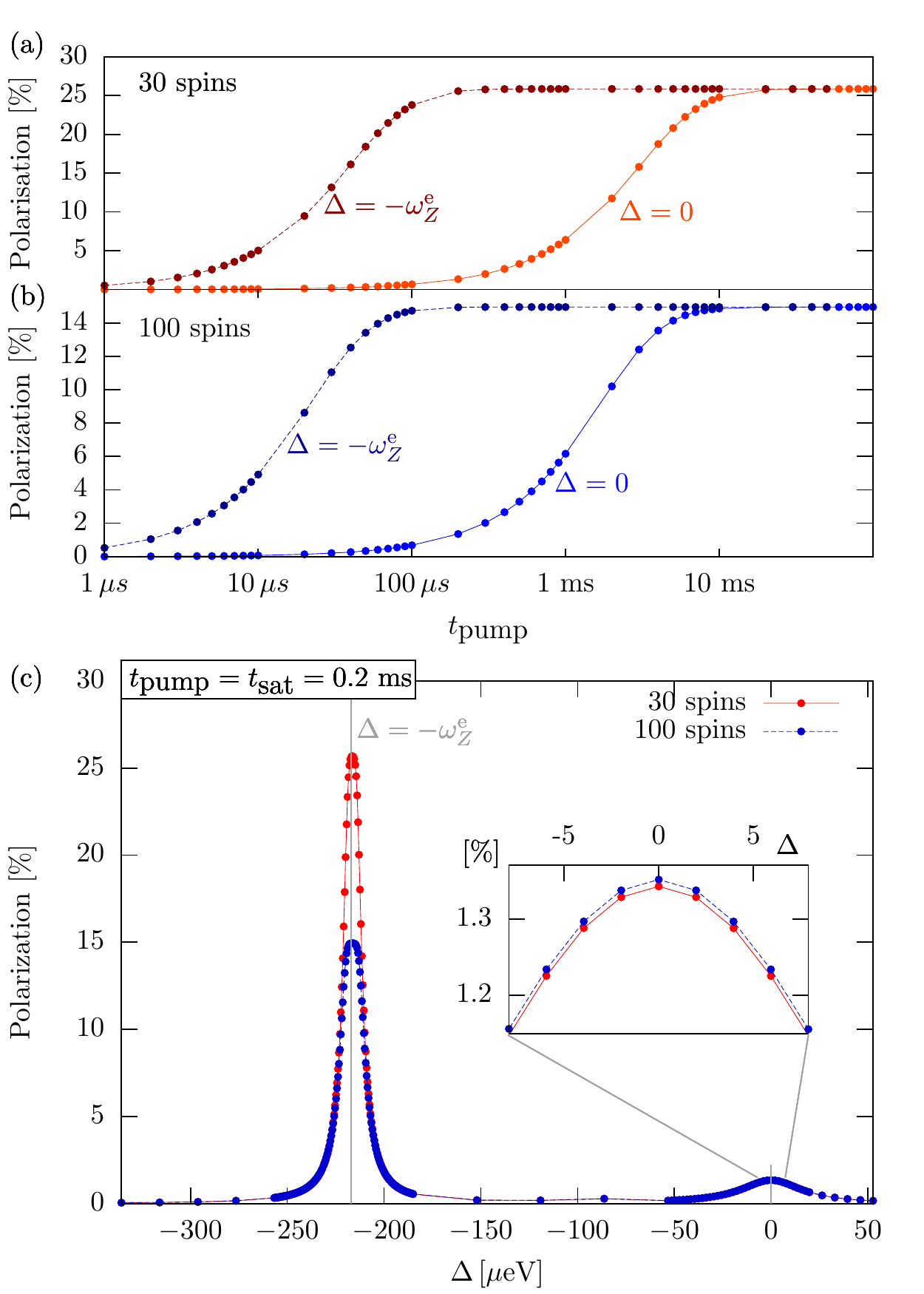}
\caption{Nuclear spin polarization calculated for the case of homogeneous hyperfine
coupling constant. (a) Saturation of the polarization for $N=30$ by pumping the
spin-forbidden transition at a laser detuning
$\Delta=-\omega_{\mm{Z}}^{\ue}=-220\,\mm{\mu eV}$ and the allowed transition at
$\Delta=0$.  (b) Same as (a), but with $N=100$.  (c) The degree of nuclear polarization
for $N=30$ and $N=100$ after a pumping time $t_{\mm{pump}}=0.2\,\mm{ms}$ as function of
laser detuning $\Delta$. The maximal degree of polarization observed for
$\Delta=-\omega_{\mm{Z}}^{\ue}$ reduces for increasing number of spins. The vertical
lines at $\Delta=-\omega_Z^\ue$ and $\Delta=0$ are visual guides to emphasize laser
dragging effects and the change of optical resonance. Inset: magnification around
$\Delta=0$ showing a difference in the polarization degree between $N=30$ and $100$.}
\label{fig:hom_result}
\end{figure}
In Fig.~\ref{fig:hom_result}, we present results for the polarization as a function of
pumping time, $t_{\mm{pump}}$, and laser detuning, $\Delta$, for different number of
spins. In order to find a percentage, we have divided $P$ by the maximally achievable
polarization, $P_{\mm{max}} = -N/2$. Here, the minus sign reflects the direction nuclear
spins are polarized with the present pumping mechanism. In accordance with the
experimental findings~\cite{chekhovich_prl2010}, we observe a build up of polarization
for $\Delta=0$ and $-\omega_{\mm{Z}}^{\ue}$, but not to the same extent [c.f.
Fig.~\ref{fig:hom_result}(c)]. Moreover, we observe a decline of the possible maximal
degree of polarization with increasing number of nuclear spins.

The fact that a polarization of 100\% is not possible for homogeneous hyperfine
coupling is attributed to the formation of a hyperfine dark state~\cite{imamoglu_prl2003,
christ_prb2007,kozlov_JETP2007,ribeiro_2009},
\begin{equation}
	\rho_{\mm{nuc}}(t \geq t_{\mm{sat}}) = \sum_j p_j (j) \ket{j -j}\bra{j -j}.
	\label{eq:homdarkstate}
\end{equation}
The population of this state cannot be changed by the hyperfine interaction, since there
is no population transfer between different $j$ blocks. 

The polarization of such a state can be straightforwardly evaluated by using $p_j (j) =
g(j)/2^N$. We have
\begin{equation}
	P^N_{\mm{sat}} = \frac{1}{P_{\mm{max}}}\sum_{j}^{\frac{N}{2}} j\, p_j(j),
	\label{eq:homPsat}
\end{equation}
where the lower bound of the sum is $j=0$ for even $N$ and $j=1/2$ for
an odd $N$.
The sum~(\ref{eq:homPsat}) can be evaluated analytically:
\begin{equation}
P_{\rm sat}^N = -\frac{1}{N}+\left\{\begin{array}{l}
\frac{2(1+2N)\Gamma\left(\frac{N+1}{2}\right)}{\sqrt{\pi} N^2
  \Gamma\left(\frac{N}{2}\right)},
\quad N \: \mm{even},\\
\frac{2 \Gamma \left(\frac{N}{2}\right)} {\sqrt{\pi}\Gamma\left(\frac{N+1}{2}\right)},
\quad N \:\mm{odd} .
\end{array}
\right.
\label{eq:exactPsat}
\end{equation}
Using Sterling's formula, we find that both expressions asymptotically 
approach $P^N \approx \sqrt{\pi/8N}$.

\begin{figure}[t!]
\centering
\includegraphics[width=1\columnwidth]{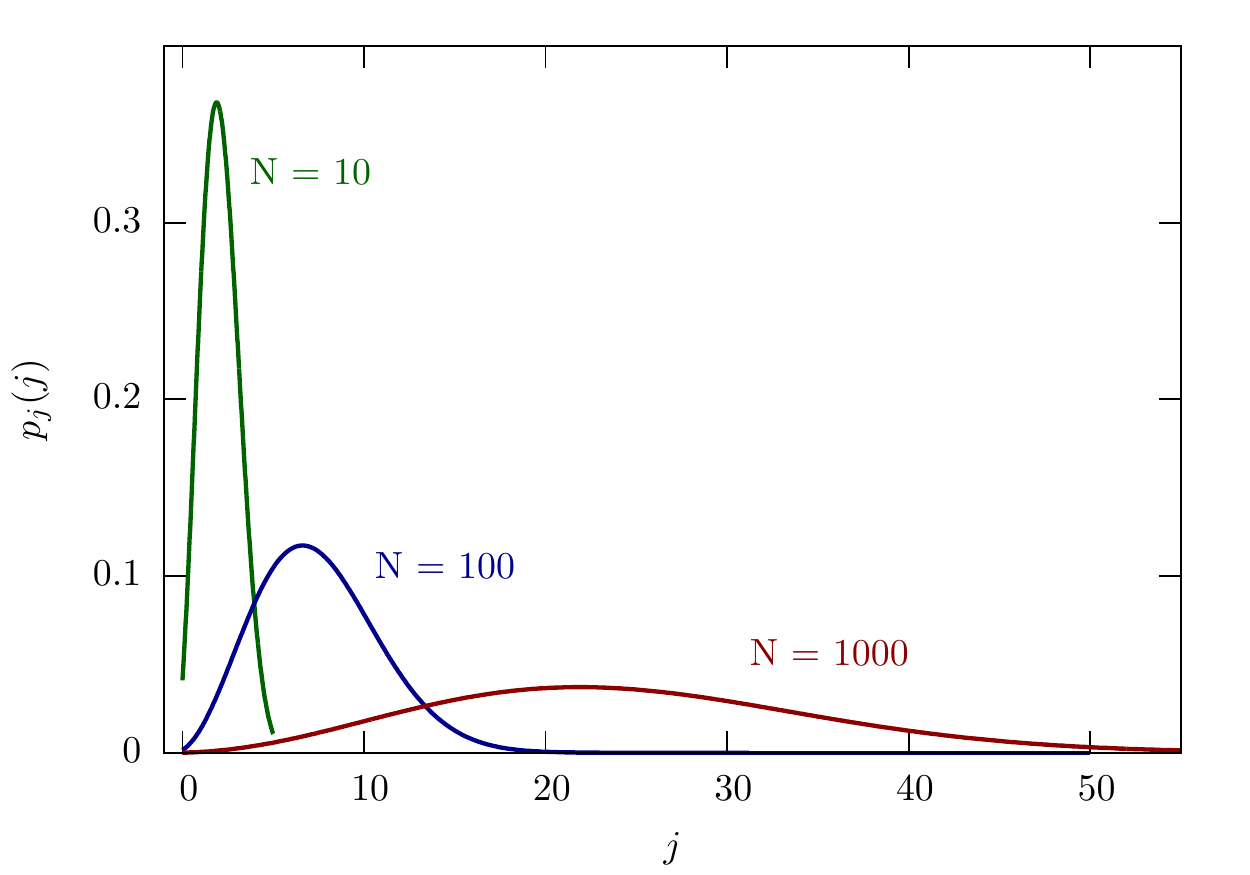}
\caption{Distribution $p_j (j)$ of the total angular momentum $j$. The larger the
system gets, the smaller the value of $p_j$ becomes for the most likely $j
\sim \sqrt{N/2}$ and the faster it converges to $0$.}
\label{fig:distj}
\end{figure}

The evaluation of Eq.~\eqref{eq:exactPsat} for $N=30$ and $100$ respectively yields,
$P^{30}_{\mm{sat}} \simeq 26.04\%$ and $P^{100}_{\mm{sat}} \simeq 14.99\%$ in a very good
agreement with our results. The asymptotic behavior can be easily understood by
considering the distribution $p_j (j)$ for increasing number of spins (cf.
Fig.~\ref{fig:distj}). For systems with a large number of nuclear spins, the distribution
only has sizable values around a small vicinity of its maximum ($j\simeq \sqrt{N/2}$).
This results in the values of $j$ that could potentially lead to high polarization, such as
$j=N/2$, to play no role in the average polarization since $p_j (N/2)_{N \gg 1} \sim 0$.  As
an example, we have $P^{10^4}_{\mm{sat}} \sim 1 \%$, with similar estimations found in
Refs.~[\onlinecite{kozlov_JETP2007}] and [\onlinecite{christ_prb2007}]. This behavior for
a large number of nuclear spins is far from experimentally observable values, therefore a
model with homogeneous hyperfine coupling cannot be used for explaining the limit of the
nuclear polarization observed in the experiments.  

We have however to point out that such a  model qualitatively reproduces all physical
phenomena observed experimentally. In addition to the already discussed similarities with
experimental data, it also reproduces dragging effects arising from the build-up of
nuclear polarization. They can be noticed in Fig.~\ref{fig:hom_result}(c), but are not
prominent, because of two reasons: we can only model a small number of nuclear spins and
achieve low degrees of polarization. Both of these facts correspond to
negligible Overhauser fields compared to the electron Zeeman splitting.  The build-up of
the Overhauser field causes the laser dragging~\cite{latta_nature2009,
chekhovich_prl2010} and changes the optical resonance conditions. This leads to the
maximal polarization to be shifted from the expected values of detuning, $\Delta=0$ and
$\Delta=-\omega_{\mm{Z}}^{\ue}$.

\subsection{Inhomogeneous hyperfine coupling}

In a more realistic model for the hyperfine interaction, an electron spin in a quantum
dot couples to nuclear spins at different lattice sites with different strengths [c.f.
Eqs.~\eqref{eq:Hhfl} and \eqref{eq:Hhft}]. In the case where the confinement is assumed
to be harmonic, we have $A_k^{\ue}=A_0^{\ue}\exp(-r_k^2/r_0^2)$~\cite{coish_prb2004,
tsyplyatyev_prl2011}, where $r_0$ is the radial size of the confinement.  $A^{\ue}_1$ is
the coupling strength of the nuclear spin in the center of the quantum dot and $r_k$ is
the radius of the $k$th shell with a constant coupling.  With nuclear spins divided
in many groups of constant coupling, the problem becomes more complex, but we can still
use the same concepts as for the homogeneous coupling case. In the latter case, the
quantum number $j$ is conserved and in the former case the conserved quantum numbers are
the total angular momenta of different groups: $j_1, j_2, \cdots, j_n$. Consequently, the
nuclear density matrix is block diagonal for different sets of $j_1, j_2, \cdots, j_n$,
and we can once more evaluate the nuclear dynamics separately for these blocks.  Since
the power of conventional computers does not allow us to compute the polarization
dynamics for many groups of spins, we consider the case of two and three groups. 

\begin{figure}[t!]
\centering
\includegraphics[width=1\columnwidth]{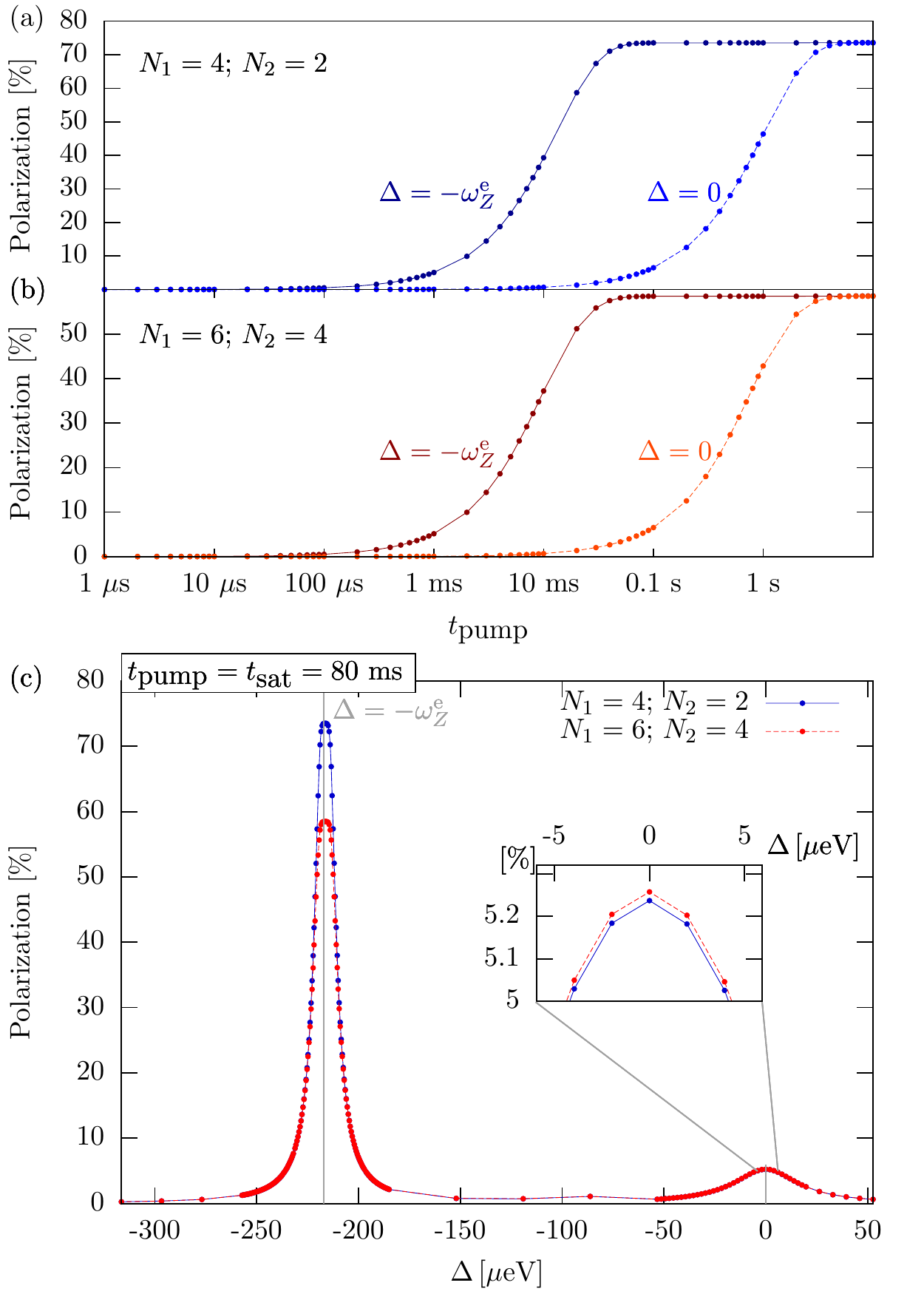}
\caption{Nuclear polarization calculated for inhomogeneous hyperfine interaction. Due
to the limitation of the computing power at our disposal, we have simulated two groups of
nuclear spins with coupling strengths $A^{\ue}_1=10^7$\,Hz and $A^{\ue}_2=10^5$\,Hz: the
saturation of the polarization for the laser detuning at the forbidden
($\Delta=-\omega_{\mm{Z}}^{\ue}$) and allowed ($\Delta=0$) transitions for (a) $N_1=4,
N_2=2$ and (b) $N_1=6, N_2=4$. (c) Nuclear polarization for different laser detunings and
for different number of spins divided into two groups for $t_{\mm{pump}}= 0.08\,\mm{s}$.
The vertical lines at $\Delta=-\omega_{\mm{Z}}^\ue$ and $\Delta=0$ are visual guides to
emphasize the shift of the optical resonance.}
\label{fig:inhom_result}
\end{figure}

In Fig.~\ref{fig:inhom_result} we present results for the nuclear polarization dynamics
for two different coupling constants. As previously, we consider a different number of
spins and compute the polarization as a function of pumping time, $t_{\mm{pump}}$, and
detuning, $\Delta$.  The results present a similar behavior as for the case of
homogeneous hyperfine coupling (cf. Fig.~\ref{fig:hom_result}). However, quantitatively
there is a difference between the reachable maximal polarizations. The saturation of the
polarization for $N_1 = 6$, $A_1^{\ue}= 10^7\,\mm{Hz}$ and $N_2 = 4$,
$A_2^{\ue}=10^5\,\mm{Hz}$ is $59\%$, it would be $42\%$ if the same number of spins were
homogeneously coupled. For $N_1=4$ and $N_2=2$, we find $74.7\%$, while it would have
been $51\%$ for six homogeneously coupled spins. 

As for the homogeneous case, the nuclear state is driven into a dark state for the
hyperfine coupling, which is a generalization of Eq.~\eqref{eq:homdarkstate}:
\begin{eqnarray}
	&&\rho_{\mm{nuc}} (t\le t_{\mm{sat}})=\nonumber\\	
	&&\sum_{j_1,\cdots,j_n}
	p_{\chi_j}(j_1,\cdots,j_n)\bigotimes_{i=1}^n \ket{j_i, -j_i}\bra{j_i, -j_i},
	\label{eq:inhdarkstate}
\end{eqnarray}
with $p_{\chi_j}(j_1,\cdots,j_n) = \prod_{i=1}^{n} g(j_i)/2^{N_i}$. We verify that this is indeed
the case by computing the degree of polarization of the nuclear state given in
Eq.~\eqref{eq:inhdarkstate}. The generalization of Eq.~\eqref{eq:homPsat} yields,
%
%
\begin{equation}
	P_{\mm{sat}}^{\{N_i\}_1^n} =
	\frac{\displaystyle \sum_{j_1\cdots j_n}^{N_1/2 \cdots N_n/2}
	(A_1^{\ue} j_1+\cdots+ A_n^{\ue} j_n)p_{\chi_j} (j_1, \cdots, j_n)}{\displaystyle \sum_{k=1}^n
	A_k^{\ue} \frac{N_k}{2}}.
	\label{eq:aklimit}
\end{equation}
Using Eq.~\eqref{eq:aklimit} we find for $N_1=6$ and $N_2=4$, $P_{\mm{sat}}^{6,4} = 59.25\%$ and
for $N_1=4$ and $N_2=2$, $P_{\mm{sat}}^{6,4} = 74.69\%$ in very good agreement with our
results.

\begin{figure}[t!]
\centering
\includegraphics[width=1\columnwidth]{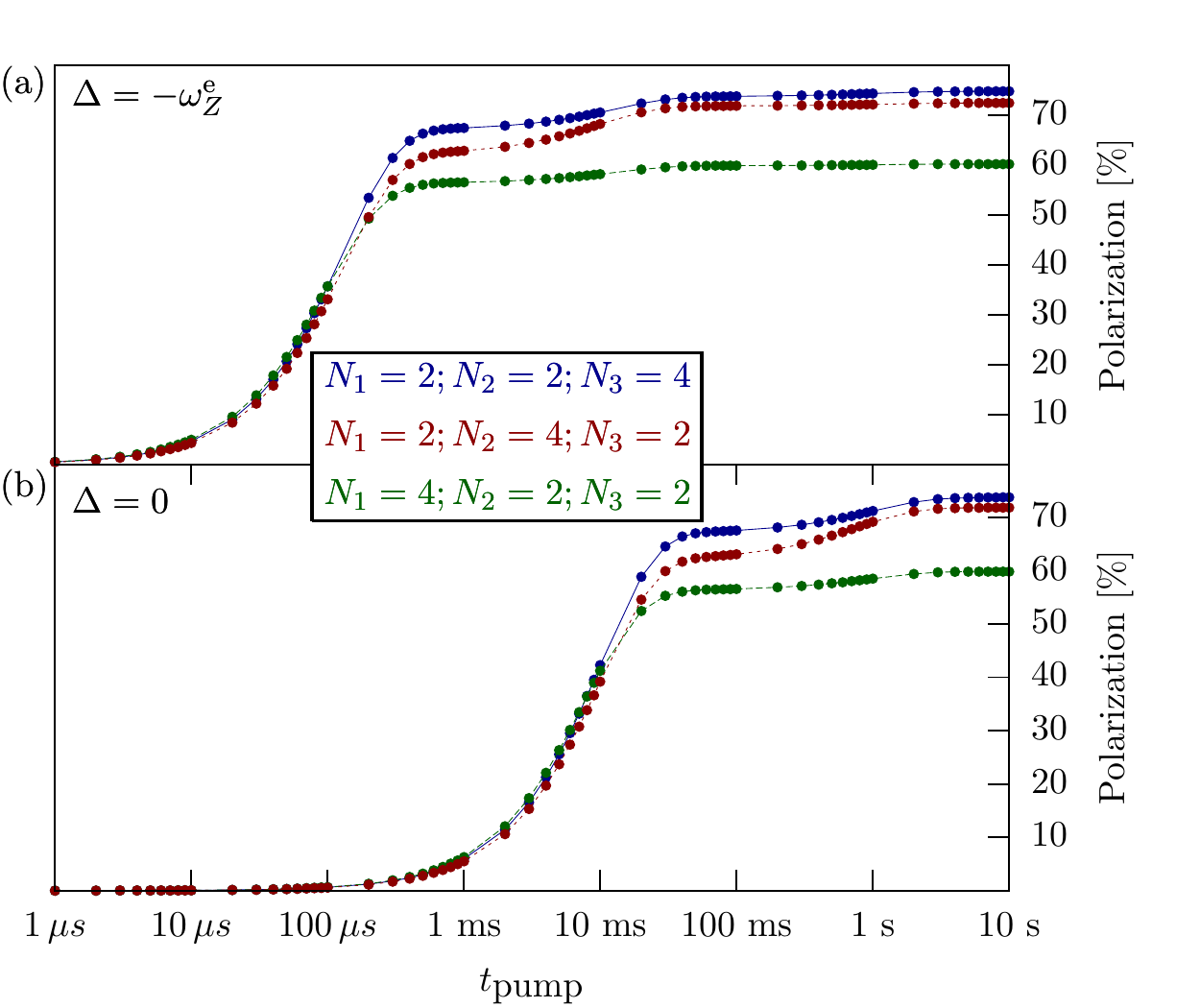}
\caption{Saturation of nuclear polarization calculated for nuclear spins divided into
three groups and inhomogeneously coupled to the electronic spin. The coupling strengths
are $A^{\ue}_1=10^8$\,Hz, $A^{\ue}_2=10^7$\,Hz and $A^{\ue}_3=10^6$\,Hz. (a) Polarization
at $\Delta=-\omega_Z^{\ue}$; and (b) at $\Delta=0$.}
\label{fig:3sectors}
\end{figure}

To confirm our observations, we have also computed the saturation of the polarization for
$n=3$. The total number of spins was kept constant, while the number of spins in the
groups was varied. We have chosen $A^{\ue}_1=10^8\,\mm{Hz}$, $A^{\ue}_2=10^7\,\mm{Hz}$,
and $A^{\ue}_3=10^6\,\mm{Hz}$. The results are presented in Fig.~\ref{fig:3sectors}.
We compare the polarization at saturation as obtained with the solution of the master
equation and calculated using Eq.~\eqref{eq:aklimit}. We find very good agreement between
both results, which indicates that the nuclear spin state is driven to the dark state
described by Eq.~\eqref{eq:inhdarkstate}.  We have found $P_{\mm{sat}}^{4,2,2} =60\%$,
$P_{\mm{sat}}^{2,4,2} = 72\%$, and $P_{\mm{sat}}^{2, 2, 4}=74.7\%$. 

As our results demonstrate, a more realistic treatment of the hyperfine interaction leads
to a substantial increase in the maximal degree of nuclear polarization. Moreover, the
results presented in Fig.~\ref{fig:3sectors} suggest that the polarization degree depends
strongly on the group configurations, i.e., the number of spins per group, strength of
the coupling (electronic envelope wave function), and (to a minor extent) on the total
number of spins.  Since Eq.~\eqref{eq:aklimit} predicts accurately the degree of
polarization, we can apply it for finding the maximal degree of
nuclear polarizations for larger systems, mimicking to some extent the nuclear spins in
quantum dots.

For a more realistic description, we assumed a system where nuclear spins form a
three-dimensional cubic lattice (InP has a zincblende lattice) and split the lattice
sites into equidistant shells from the maximum of the electron wave function, which is
assumed to have a Gaussian distribution. For a system of $360$ spins we obtained 25.5\%
of polarization and for $365$ spins 30.4\%. The difference between the two cases comes
from the relative position of the maximum of the electron wave function relative to the
lattice: in the first case, it was set in the middle of the unit cell, and in the second
case, it was exactly at a lattice site. This result clearly indicates that there is
no universal value for $P_{\mm{sat}}$, but rather that every quantum dot has a different
saturation polarization. We consider now a quantum dot made out of two elements (two
sublattices) with different hyperfine coupling constants. We have to group the nuclear
spins not only by considering the distance to the center, but also by taking into account
the different hyperfine couplings of each species. If we consider the previous cases, but
we divide the lattice into two sublattices, the degree of nuclear spin polarization of
the dark state becomes~34.4\% and 40.6\%, respectively. This calculation shows that in realistic
quantum dots, which consist of two and very often more elements, the achievable degree of
polarization is higher than the one calculated with identical nuclear spins. We show in
Appendices~\ref{sec:appendix1} and \ref{sec:appendix2} the explicit distribution of $A_k$'s
and $N_k$'s for both considered cases.

In our calculations, we have considered systems consisting of nuclear spins with $I=1/2$,
whereas the relevant optically active semiconductor quantum dots are built up by
materials (e.g., In and P) with $I\geq 1/2$. We can therefore raise the question if some
of the neglected interactions could significantly affect the nuclear spin pumping. In our
model, we have neglected nuclear-nuclear spin interactions, i.e., nuclear Zeeman energy,
dipole-dipole coupling, and quadrupole splittings, the latter only being relevant for
$I>1/2$. Among these, nuclear dipole-dipole interaction constitutes a competing mechanism
that could prevent the formation of the dark state. However, experimental findings
indicate that nuclear spin diffusion happens on time scales ranging from seconds to
hours~\cite{coish_statsolidi2009,tartakovskiiprivate}. This indicates a small dipole
coupling that can be neglected in comparison with the hyperfine-mediated nuclear
dipole-dipole coupling, and which is the main source of diffusion during the pumping
cycle. The effect of the nuclear Zeeman and quadrupole splitting is more subtle. The
obvious change concerns Eq.~\eqref{eq:fr_rate}, where the denominator would also include
the difference in nuclear Zeeman energy and quadrupole splitting between the nuclear
states $\ket{\chi}$ and $\ket{\chi'}$. These are small corrections in comparison with the
electron Zeeman energy and therefore they do not alter Eq.~\eqref{eq:fr_rate}
significantly. However, when considering different isotopes, the nuclear Zeeman and
quadrupole interactions force an additional division of nuclear spins with the same
distance from the maximum of the electron wave function. Different isotopes have
different nuclear gyromagnetic ratios and quadrupolar splittings which lead to slightly
different forbidden relaxation rates. Such additional fragmentation leads to an increase
of the maximal degree of polarization. A smaller number of nuclear spins per shell
leads to an increase of the statistical weight of the states contributing the most to the
degree of polarization (cf. Fig.~\ref{fig:distj}).

\section{Conclusions}

We have developed a master-equation formalism that allows us to partially explain recent
experimental observations~\cite{chekhovich_prl2011} on the saturation of the nuclear spin
polarization when pumped via an optically spin-forbidden transition between a heavy hole
and a trion state. We have identified both mechanisms leading to spin polarization
depending on the laser detuning and we have found flip-flop rates that depend explicitly
on the nuclear state. 

Based on our formalism, we have calculated the exact time evolution of the nuclear spin
polarization. Considering the nuclear spin bath as an ensemble of quantum spins, which is
initially in thermal equilibrium, we have investigated two possible models for the
hyperfine interaction: homogeneous and inhomogeneous. In both cases, the saturation of
the nuclear polarization is attributed to the conservation of the total angular momentum
of the whole nuclear state in the homogeneous case or of the particular groups of same
coupling for the inhomogeneous case. In the latter, the degree of maximal nuclear
polarization is consistent with the experimentally observed values. Our findings show
that variations in the maximal degree of polarization depend on the chemical composition
of the quantum dot and the distribution of the electron wave function inside of the
quantum dot. However, the latter property offers a possible way to overcome the limit set
by the dark state. It is possible to change the functional form of the electronic wave
function by applying electric fields~\cite{imamoglu_prl2003,christ_prb2007}. This would
lead to a redefinition of the nuclear spin groups with the same hyperfine coupling constant,
which would further allow hyperfine-mediated nuclear spin pumping. It has still to be
proven experimentally that such a protocol can indeed achieve higher degrees of nuclear
spin polarization than the one set by a dark state.

\section{Acknowledgments}

We acknowledge funding from the DFG within SFB 767 and SPP 1285 and from BMBF under the program
QuaHL-Rep. H. R. also acknowledges funding from the Swiss NF.

\global\long\def\theequation{A.\arabic{equation}}

\global\long\def\thefigure{A.\arabic{figure}}

\setcounter{equation}{0}

\setcounter{figure}{0}

\appendix

\section{$A_k$ and $N_k$ distributions for a system of $360$ nuclear spins}
\label{sec:appendix1}

\begin{figure*}[h!]
\centering
\includegraphics[width=1.7\columnwidth]{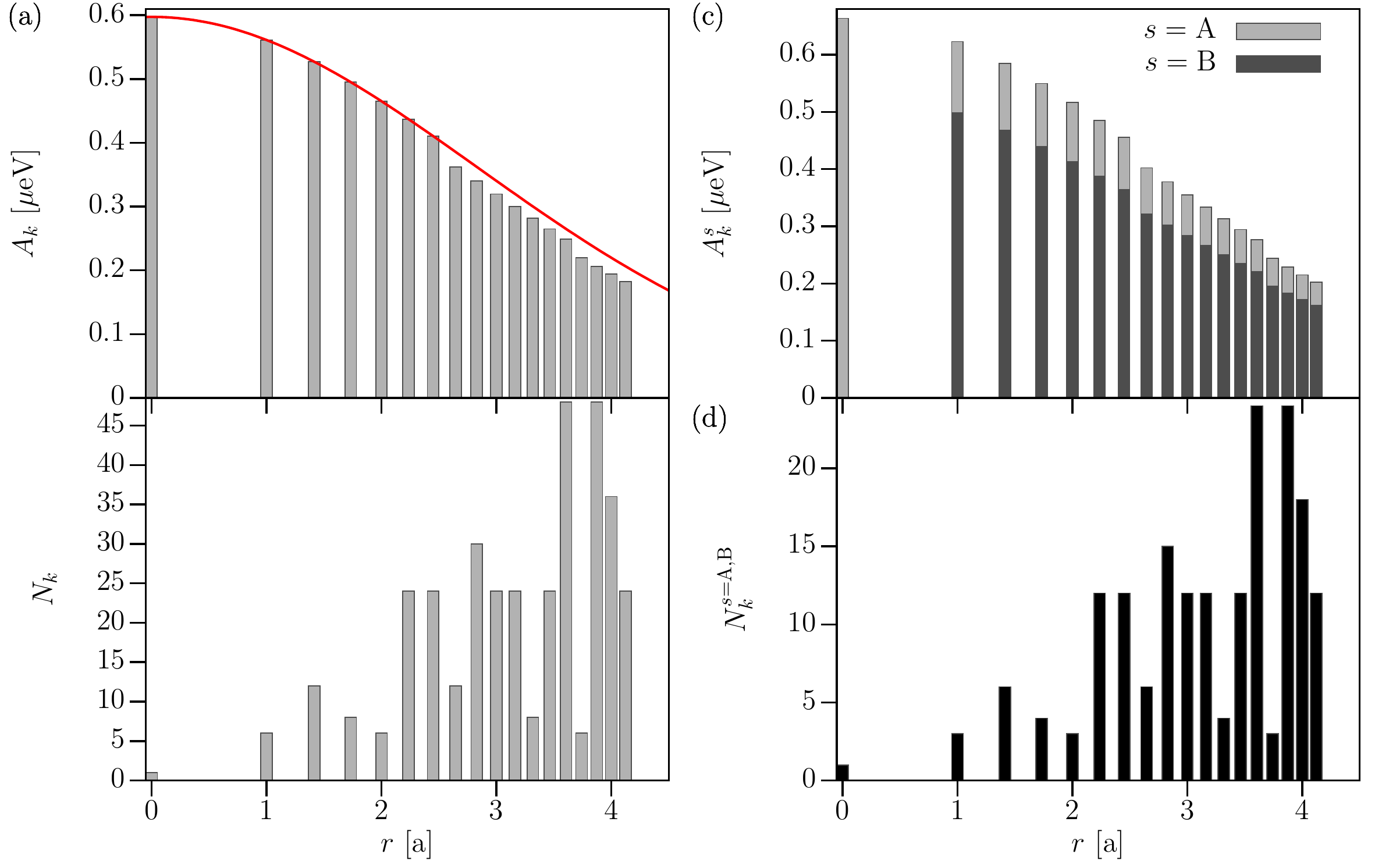}
\caption{(a) Hyperfine coupling constants $A_k$ and (b) number of nuclear spins $N_k$ as
a function of the distance to the maximum of the wave function. The distance is given in
units of the lattice constant $\rm{a}$. We have plotted in red the continuous
distribution of hyperfine coupling constants $A(\rv)=A_0 \exp[-(r/r_0)^2]$. (c) Same as
(a) for two sublattices $s=\rm{A,\,B}$. We denote the hyperfine coupling constant by
$A_k^s$. (d) Number of nuclear spins $N_k^s$ in each sublattice.}
\label{fig:appendix1}
\end{figure*}

In Fig.~\ref{fig:appendix1}, we present the strength of the hyperfine coupling constants
$A_k$ [Fig.~\ref{fig:appendix1}(a)] and the number of nuclear spins $N_k$
[Fig.~\ref{fig:appendix1}(b)] as a function of the distance to the maximum of the wave
function for a system made of identical nuclear spins. The distance is measured in
units of the lattice constant $\rm{a}$. We note that the maximum of the wave function is
not situated at a lattice site. For a three-dimensional cubic lattice, the $360$
nuclear spins are divided among 10 groups. 

In Fig.~\ref{fig:appendix1}(c), we show the distribution of $A_k^s$ and in
Fig.~\ref{fig:appendix1}(d) the number of nuclear spins $N_k^s$
($s=\rm{A,\,B}$) for a system with two nuclear species. The nuclear spins of each species
form a sublattice denoted $\rm{A}$ and $\rm{B}$. We have assumed for simplicity that each
sublattice is constituted by the same number of nuclear spins.

\section{$A_k$ and $N_k$ distribution for a system of $365$ nuclear spins}
\label{sec:appendix2}

\begin{figure*}[h!]
\centering
\includegraphics[width=1.7\columnwidth]{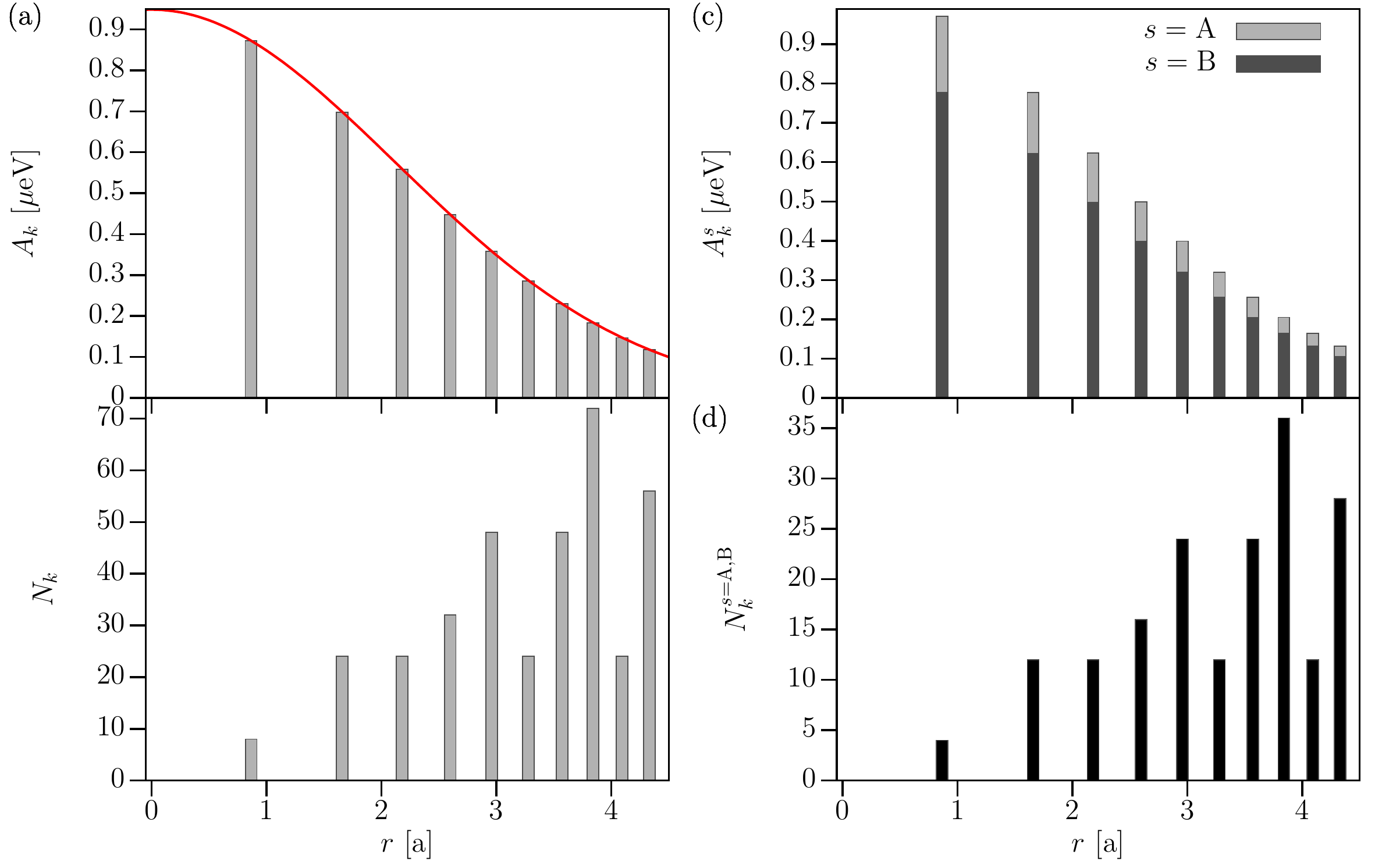}
\caption{(a) Hyperfine coupling constants $A_k$ and (b) number of nuclear spins $N_k$ as
a function of the distance to the maximum of the wave function, which is located at a
lattice site. The distance is measured in
units of the lattice constant $\rm{a}$. The continuous
distribution of hyperfine coupling constants, $A(\rv)=A_0 \exp[-(r/r_0)^2]$, is shown in
red. (c) Same as (a) for a system made of two nuclear species divided into two
sublattices ($s=\rm{A,\,B}$). (d) Number of nuclear spins $N_k^s$ for $s=\rm{A,\,B}$.}
\label{fig:appendix2}
\end{figure*}

In Fig.~\ref{fig:appendix2}, we present the same data as in the previous appendix, but for the maximum of
the wave function situated at a lattice site. 
\footnotetext{Current address:}

\end{document}